\documentclass[aps,pra,twocolumn,superscriptaddress,floatfix]{revtex4}
\usepackage{epsfig,color,graphicx}
\usepackage{amsmath,amssymb,amsfonts,amsthm}
\begin{document}

\title{Quantum distillation: dynamical generation of low-entropy states of strongly 
correlated fermions in an optical lattice}

\author{F. Heidrich-Meisner} 
\affiliation{Institut f\"ur Theoretische Physik C, RWTH Aachen University, 52056 Aachen, Germany}
\author{S. R. Manmana}
\affiliation{Institute of Theoretical Physics, \'Ecole Polytechnique F\'ed\'erale de Lausanne, CH-1015 Lausanne, Switzerland}
\author{M. Rigol}
\affiliation{Department of Physics, Georgetown University, Washington, District of Columbia 20057, USA}
\author{A. Muramatsu}
\affiliation{Institut f\"ur Theoretische Physik III, Universit\"at Stuttgart, 70550 Stuttgart, Germany}
\author{A. E. Feiguin} 
\affiliation{Department of Physics and Astronomy, University of Wyoming, Laramie, WY 82071, USA}
\affiliation{Microsoft Project Q, University of California, Santa Barbara, CA 93106, USA} 
\affiliation{Condensed Matter Theory Center, University of Maryland, College Park, MD 20742, USA}
\author{E. Dagotto}
\affiliation{Materials  Science and Technology Division, Oak Ridge National Laboratory,
 Oak Ridge, Tennessee 37831, USA and\\
 Department of Physics and Astronomy, University of Tennessee, Knoxville,
 Tennessee 37996, USA}

\date{October 19, 2009}

\begin{abstract}
Correlations between particles can lead to subtle and sometimes
counterintuitive phenomena. We analyze one such case, occurring during the
sudden expansion of fermions in a lattice when the initial state has a strong
admixture of double occupancies. We promote the notion of quantum
distillation: during the expansion, and in the case of strongly repulsive
interactions, doublons group together, forming a nearly ideal band
insulator, which is metastable with a low entropy. We propose that
this effect could be used for cooling purposes in experiments with
two-component Fermi gases.
\end{abstract}

\maketitle

One of the most exciting features about ultracold atom gases is the possibility
of experimentally studying strongly correlated systems with time-dependent 
interactions and  in out-of-equilibrium situations 
(for a review, see \cite{bloch08}). For instance, a time-dependent 
tuning of parameters has been used to illustrate the collapse and revival of 
coherence properties of bosons in an optical lattice suggested to be well 
described by the Bose-Hubbard model \cite{greiner02}. Also, the question of 
the relaxation of strongly correlated systems initially prepared in a high-energy 
state that typically is not an eigenstate has been studied \cite{trotzky08}, 
as well as the intimately related question of thermalization \cite{kinoshita06}.
A third context is the expansion of originally trapped particles into an empty 
optical lattice or a waveguide.

While expansions after turning off all trapping potentials and lattices are 
commonly used to measure the momentum distribution of the originally trapped system 
\cite{bloch08}, recently, several studies have investigated 
the expansion  in one dimensional (1D) geometries with and without the presence of 
an optical lattice along the 1D tubes. In 1D cases, in contrast to 
the usual time-of-flight experiments, interactions play a fundamental role. The findings 
include  phenomena such as the fermionization of the momentum 
distribution function of hard-core bosons and anyons \cite{fermionization}, 
the asymptotic transformation of generic Lieb-Liniger wave-functions to a 
Tonks-Girardeau structure \cite{buljan}, the emergence of quasi-coherence in 
bosonic \cite{hcbs} and fermionic systems \cite{hm08a}, or disorder-induced 
effects \cite{disorder-exp}. Such 
expansions have already experimentally been realized in 1D ultracold atomic gases \cite{exp-expansion}.

While  strongly correlated phases of bosons in  optical lattices 
have been observed and studied in many experiments, 
reaching the quantum degenerate regime of fermions is more 
difficult  \cite{bloch08,ketterle}. Only recently, evidence for the observation 
of the Mott-insulator (MI) transition of a two-component Fermi gas in optical lattices has been 
presented \cite{fermions-exp}. Unfortunately, the temperatures in those experiments are
still too high to observe the expected low-temperature insulating antiferromagnetic phase, 
and to make contact with the regime of interest to high-$T_c$ superconductivity \cite{hofstetter02}.
Hence, intense efforts are currently under way to develop efficient cooling schemes for 
fermions in optical lattices \cite{fcooling}.

In this Rapid Communication, we study the expansion of a two-component Fermi gas in an 
optical lattice from an initial state with a strong admixture of doubly occupied sites. 
Our main result is the observation that double occupancies (doublons)  are dynamically separated from the rest of the
system and that they group together into a metastable state, which is very close to a Fock 
state. The condition for this to happen is that interaction energies need to be much larger than the 
kinetic energy.  Our results are in qualitative agreement with recent studies 
that
have argued, invoking a similar reasoning,  that relaxation times can be rather long, and one may thus  encounter metastable 
states (see, {\it e.g.}, Ref.~\cite{rosch08}). Experimentally, this phenomenon has been 
observed as repulsively bound pairs in the case of bosons \cite{winkler06}.
We will also discuss the potential of our findings as a cooling technique where 
one dynamically generates a low-entropy region in a system with arbitrarily large 
values of the on-site repulsion.

{\it Model and methods --} We consider the one-dimensional (1D) Hubbard model:
\begin{equation}
H_0 = -t\sum_{i=1}^{L-1} ( c_{i+1,\sigma}^{\dagger}  c_{i,\sigma}^{ } 
+ \mathrm{h.c.} ) + U \sum_{i=1}^{L}
n_{i,\uparrow} n_{i,\downarrow} \,. \label{eq:ham}
\end{equation}
Standard definitions are employed in Eq.~(\ref{eq:ham}) \cite{hm08a}. Open boundary 
conditions are imposed, $L$ is the number of sites and $N$ the number of particles 
(with a filling factor $n=N/L$). The nonequilibrium dynamics is studied using 
the adaptive time-dependent density matrix renormalization group method (tDMRG) \cite{tDMRG}
and time-dependent exact diagonalization (ED) techniques \cite{manmana05}.

We prepare initial states with particles confined into a finite region of an optical 
lattice, with a filling factor of $1<n_{\mathrm{init}}<2$  in that 
region and zero otherwise. To this end \cite{hm08a}, we apply   onsite energies $\epsilon_i$ to only a 
portion of the system: $H_{\mathrm{conf}}=\sum_{i=1}^{L} \epsilon_i n_i$
($\epsilon_i \sim 10^{6}t$ for
$1<i_0<i\leq L$ and $\epsilon_i=0$ elsewhere). Hence, 
at time $\tau\leq 0$, we have $H= H_0+H_{\mathrm{conf}}$,  
 while 
we turn off $H_{\mathrm{conf}}$ at $\tau=0^+$. In our tDMRG runs, we use either a third-order 
Trotter-Suzuki time-evolution scheme  or a Krylov-space 
based method \cite{manmana05}, with time steps of $\delta\tau\,t=0.01$ and $0.005$ (we set $\hbar=1$). During the time-evolution, 
we keep up to 1600 states. 

\begin{figure}[t]
\includegraphics[width=0.38\textwidth]{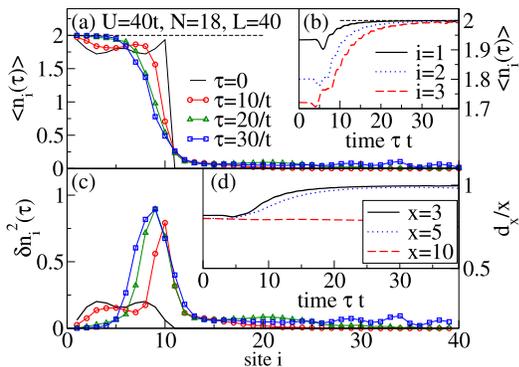}
\caption{(Color online) Expansion from an initial state with 
$n_{\mathrm{init}}=1.8$ ($N=18$, $U=40t$): (a) Density $\langle n_i(\tau)\rangle$ and 
(c) local charge fluctuations $\delta n_i^2$ for times $\tau \, t =0,10,20, 30$. Insets: 
(b): $\langle n_i(\tau)\rangle$ vs time for $i=1,2,3$. (d) Average double occupancy 
$d_{x}(\tau)/x$  for $x=3,5,10$ ($d_{x}(\tau)=\sum_{i=0}^{x} \langle \tilde d_i(\tau)\rangle$).}
\label{fig:ni}
\end{figure}

{\it Results --} Figure~\ref{fig:ni}(a) shows snapshots of density profiles $\langle n_i(\tau) \rangle$ 
at several times for $U=40t$ and $n_{\mathrm{init}}=1.8$ (for results for other $n_{\mathrm{init}}$, see  \cite{epaps}). Since $U\gg 4t$, single-particle 
hopping can only propagate a particle out of a doubly occupied site into the empty sites 
to the right of the initially occupied region if, at the same time, the remaining particle 
is moved to the left into a site that was previously singly occupied, in order to preserve 
the total energy. For that reason, the density in the first sites 
{\it increases} as the remaining of the block slowly ``melts''. From Fig.~\ref{fig:ni}(b), 
one can see that the density in the first sites gets very close to $\langle n_i\rangle=2$, 
which promotes the picture that dynamically, doublons are separated from the 
rest and group together in a region of the lattice. 

We introduce the term quantum distillation for this process. Figure~\ref{fig:ni}(c) 
shows the local charge fluctuations $\delta n_i^2=\langle n_i^2\rangle-\langle n_i\rangle^2$. 
Consistent with the picture described before, these fluctuations are the largest 
in the interface region (sites 7-11 in the figure), while they die out in the first 
sites as time increases. To gain a better understanding of how this happens with time, we 
compute the average double occupancy $d_x/x$ ($d_x=\sum_{i=0}^{x} \langle \tilde d_i(\tau)\rangle$; 
$\tilde d_i=n_{i\uparrow}n_{i\downarrow}$) on the $x$ first sites, counting from the left. 
Figure~\ref{fig:ni}(d) depicts this quantity for different values of $x$. For $x<10$ and, 
{\it e.g.}, $x=3,5$, the average double occupancy 
increases towards $d_x/x=1$, which is quite stable over a long time window for 
$U\geq 40t$ [Fig.~\ref{fig:ni}(d)]. For $x=10$, {\it i.e.}, the block with $\langle n_i\rangle >0$ at $\tau=0$,
 this quantity remains almost constant: this is basically 
equivalent to saying that  only a small portion of  the interaction energy   
$E_{\mathrm{int}}=U\sum_{i=1}^{10} \langle \tilde d_i\rangle$ is  converted into kinetic energy 
during the times simulated. 

\begin{figure}[t]
\includegraphics[width=0.38\textwidth]{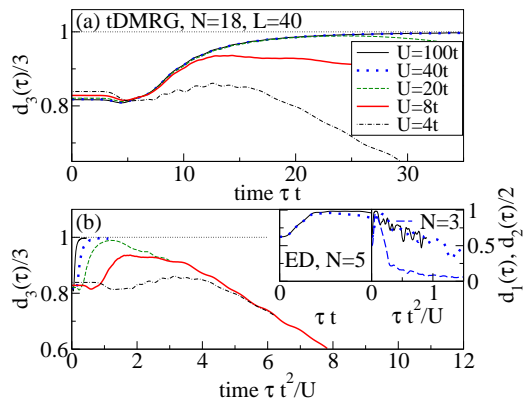}
\caption{(Color online)
Average double occupancy $d_{3}(\tau)/3$  ($n_{\mathrm{init}}=1.8$): vs (a) time $\tau$
and (b) vs $\tau\, t^2/U$ for $U/t=4,8,20,40,100$ (tDMRG). Insets in (b): $d_1$ 
for $N=5$ ($U/t=40, 100$); $d_2/2$ for $N=3$ ($U=40t$, thin dashed line) (ED). }
\label{fig:double}
\end{figure}

{\it Time scales --} 
We next address the question on which time scales (i) the Fock state 
is formed and (ii) it decays.
 The latter is equivalent to asking on what  time 
scale a band insulator (BI) in the large $U$ limit would decay. 
While from energy considerations it is clear that a piece of doubly occupied sites, 
{\it i.e.}, a BI,
delocalizes on a time-scale of  $U/t^2$ \cite{winkler06}, we are primarily interested in  the case
of $1<n_{\mathrm{init}}<2$.
To address this situation,
we first perform ED calculations for a short initial state with $1<n_{\mathrm{init}}<2$, 
specifically $N=3,5$ on two and  three sites, respectively [see the insets of Fig.~\ref{fig:double}(b)]. In this case, 
one can clearly see that doublons form on the first sites on a $U$-independent time-scale, 
governed by the bare hopping matrix element $t$. Doublons then 
get delocalized over the neighboring empty sites on a time-scale $U/t^2$ (times a prefactor that scales with $N$), consistent with the 
decay-rate of a BI.

Our tDMRG results presented in Figs.~\ref{fig:double}(a) and (b) unveil that the same picture holds
 for $n=1.8$ with $N=18$. Figure~\ref{fig:double}(a) shows the average 
double occupancy on the first three sites for several values of $U$. 
Quantum distillation is best realized for $U > 8t$ (as opposed
to $U \lesssim 8t$), since, at $U=8t$, the average
 double occupancy 
increases with time ($\tau>5/t$), but it reaches a maximum value that is well below one. 
Nevertheless, as the large-$U$ results show, the formation of a quasi-Fock state happens 
over a $U$-independent time-scale. Figure~\ref{fig:double}(b) displays the same data 
versus $\tau\,t^2/U$, and the small-$U$ curves fall on top of each other at times $\tau\, t^2/U>4$, 
while the large-$U$ data seem to approach  the same curve, too. 
The results of Fig.~\ref{fig:double}(b) show, in the example of $x=3$, that the Fock state will start to delocalize after a time $\tau \sim 50/t$ in the $U=40t$ case ($\tau\sim 120/t$ for 
$U=100t$). 

While we have presented results for $N=18$,
the physics does not qualitatively change upon increasing $N$ at a fixed $n_{\mathrm{init}}$ \cite{epaps}. 
Essentially, the time scale 
for building up the Fock state scales linearly with $N$ \cite{epaps}.
Note that in typical experiments with 1D optical lattices, on the order of 100 atoms are confined in one 1D structure \cite{bloch08}.
It should be readily possible to experimentally observe the quantum distillation by measuring, {\it e.g.}, the radius $R_d$
of the double occupancies [$R_d^2 (\tau)\propto \sum_i \langle \tilde d_i(\tau)\rangle  i^2 $, see \cite{epaps}].
We predict that, for $U\gg 4t$ and $n_{\mathrm{init}}>1$, this quantity  should decrease as a function of time. 
As $U/t$ depends exponentially on the lattice depth $V_0$, one can enter into the regime $U\gg 4t$ by tuning $V_0$ \cite{quench}. 

{\it Entanglement entropy -- } Further quantitative information can be gained by invoking  the entanglement entropy 
$S_x=- \mbox{tr}\lbrack \rho \ln(\rho)\rbrack$ (Ref.~\cite{statmech}), where $\rho$ is the reduced density matrix 
of a subsystem of length $x$ \cite{schollwoeck05}. The subsystems of primary interest here are those
where double occupancy increases as time evolves, as illustrated in Fig.~\ref{fig:ni}(d) 
($x$ counts the sites starting from the left end). 

\begin{figure}[t]
\includegraphics[width=0.38\textwidth]{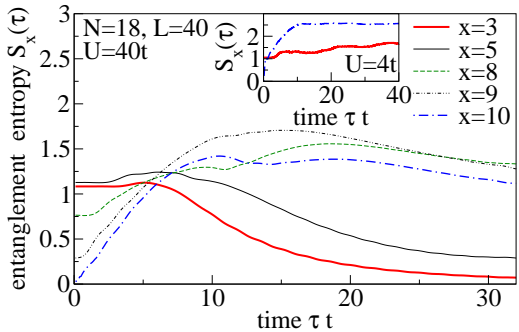}
\caption{
(Color online) 
 Entanglement entropy $S_x$ ($n_{\mathrm{init}}=1.8$, $U=40t$). 
 Inset: $S_x$ for $x=3,10$ at $U=4t$.
 }
\label{fig:entropy}
\end{figure}

Figure \ref{fig:entropy}  depicts one of our most important results, constituting a
defining property of the quantum distillation process, namely, the spontaneous reduction 
of the entanglement entropy in the metastable region where doublons group together. 
We plot $S_{x}$ for several $x\leq10$
at $U=40t$, which ought to be contrasted against the corresponding $U=4t$ data, shown in the inset. Figure \ref{fig:entropy} shows that for the leftmost sites,  $S_x(\tau=0)$ is
nonzero and remains approximately constant for some $x$-dependent time
 window ({\it e.g.}, $\tau t\lesssim 5$ for $x=3$).
At later times, the behavior 
strongly depends on the value of $U/t$. For $U/t=4$, 
$S_3$ increases as the 
density decreases, while for $U/t=40$, $S_{3}$ decreases, on the time scales simulated, by a factor of 15
as the metastable state with $\langle n_i\rangle=2$ is generated. We relate this decrease of $S_x$
to the quantum distillation, which creates 
regions with low entanglement. This is  complementary to experiments aiming at preparing 
maximally entangled states \cite{kwiat01}.

{\it Harmonic trap --} While there are experimental efforts directed at engineering box-like traps (see, {\it e.g.}, Ref.~\cite{meyrath05}), which is the case our results
presented so far directly apply to, in most  typical experiments with optical lattices, 
the atoms experience a harmonic confinement \cite{bloch08}. We now consider this
case to show that quantum distillation works even in the more difficult case of strongly inhomogeneous initial states,
 using $H_{\mathrm{conf}}=V_{\mathrm{trap}}\sum_{i=1}^L (i-1)^2 n_i$,
where $V_{\mathrm{trap}}$ is the curvature of the harmonic potential.
The confinement gives rise to a shell structure in the fermionic density \cite{rigol04}: 
metallic and MI shells alternate, depending on the characteristic density 
$\tilde{\rho}=N \sqrt{V_{\mathrm{trap}}/t}$ \cite{rigol04}. 
In order to achieve optimal distillation and thus the formation of a low-entropy region, 
one does not need  to follow  the approach often discussed in the literature (see, {\it e.g.}, Ref.~\cite{fcooling}), in 
which $U/t$ is increased adiabatically. Instead, we propose that one can start with 
a trapped system with a low value of $U/t=2$. 
In such cases, double occupancy is energetically favorable against trapping energy 
and hence ideal for our distillation scheme, and moreover, the presence of MI shells is suppressed. 
 At time $\tau=0$, one can then quench 
$U/t$ to a large value \cite{quench} and turn off the trapping potential, but keeping 
the optical lattice on. In many experiments, the trapping potential is provided by 
the same lasers that produce the lattice, which, however, is not necessary. 
In order for our scheme to work, one
needs a trapping potential that can be controlled independently of the lattice 
\cite{exp-expansion}.

\begin{figure}[t]
\includegraphics[width=0.38\textwidth]{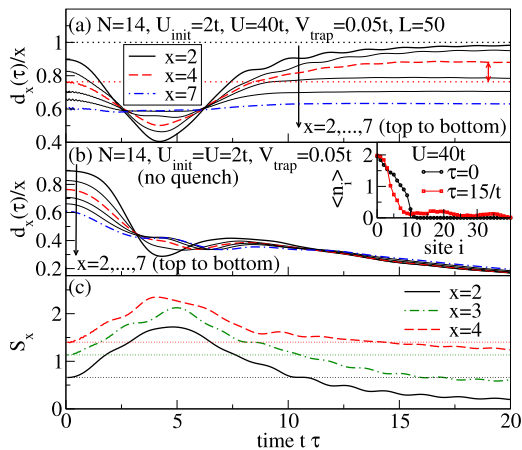}
\caption{
(Color online) Average double occupancy $d_x/x$ during the expansion from a harmonic 
trap ($V_{\mathrm{trap}}=0.05t$, $U_{\mathrm{init}}=2t$, $N=14$, $x=2,3,\dots, 7$) with (a) a quench to $U=40t$ and (b) no quench. Dotted line in (a): $d_{x=4}(\tau=0)$. 
Inset in (b): density profile at $\tau\,t=0,15$. (c) Entanglement entropy $S_x$ for $x=2,3,4$ 
($U=40t$); horizontal lines are $S_x(\tau=0)$.  }
\label{fig:trap}
\end{figure}

The tDMRG results for the time-evolution of the quench described above are displayed in 
Fig.~\ref{fig:trap}. We have taken the final values of $U/t$ after the quench to be 
$U=40t$ [Fig.~\ref{fig:trap}(a)] and $U=U_{\mathrm{init}}$ [Fig.~\ref{fig:trap}(b)]. Both
figures show the average double occupancy $d_x(\tau)/x$ for several values of $x$. 
In Fig.~\ref{fig:trap}(a), one can see that $d_x(\tau)/x$ first decreases 
and then increases to form the quasi-Fock state. 
It is also illustrative to compare 
the initial density profile to one at later times [inset of Fig.~\ref{fig:trap}(b)]: 
while fast particles escape, a block of particles remains in the initial region of the system.
A very different behavior of $d_x$ can be 
seen in Fig.~\ref{fig:trap}(b). There, since the final value of $U/t$ is small, the system
simply melts and no signature of quantum distillation is seen. 

During the expansion, the entanglement growth \cite{entanglement} induced by the quench from $U=2t$ to $40t$ 
competes with the reduction of the entanglement  entropy $S_x$ of blocks with an increasing double occupancy due to the quantum distillation.
It is thus an amazing result that, at sufficiently long times and despite the large final value of $U$ chosen in the quench,
the quantum distillation wins, as is shown 
in Fig.~\ref{fig:trap}(c). 
There, we display $S_x$
for $x=2,3,4$, and indeed,  these  $S_x$ first increase due to the quench, but eventually drop below their
initial value (dotted lines) at the maximum times simulated. For the time-scales  considered in Fig.~\ref{fig:trap}(c), we find that   $S_2(\tau) \lesssim 0.25 S_2(\tau=0)$. The same behavior emerges for other values of
$\tilde\rho$, $U_{\mathrm{init}}$, and $U$ \cite{epaps}. 

So far, our results suggest that quantum distillation is robust against a
variation of initial conditions and quenches. We envision it as a way to
experimentally achieve very low temperatures in fermionic systems by creating
a very low entropy band insulating state with arbitrarily large values of
$U/t$. Once such a state is created, it can be used as an initial state to
achieve an antiferromagnetic MI with one particle per site. The
idea would be to load such a BI in a trap for which that state is
close to the ground state. This step is important in ensuring a low-entropy initial state that is in equilibrium \cite{comm_weiss}.
 One can then adiabatically reduce the strength of that trapping
potential and of the ratio $U/t$. That way one could produce a final low
temperature Mott insulating state starting from a BI.

In conclusion, we demonstrated that low-entropy states of two-component Fermi gases 
can be dynamically created in optical lattices utilizing the expansion of particles into 
the empty lattice in the limit of strong interactions.  After having the low entropy state in an appropriate trap, an
adiabatic lowering of trapping potential and interaction strength can
lead to a final low temperature MI.
We stress that the quantum distillation we discussed here 
for fermions will also work in the bosonic case.

\begin{acknowledgments}
We thank L. Hackerm\"uller, G. Refael, A. Rosch, U. Schneider, and D. S. Weiss for 
fruitful discussions. M.R. was supported by startup funds from Georgetown University 
and by the US Office of Naval Research. A.M. acknowledges partial support by the DFG 
through SFB/TRR21. E.D. was supported in part by the NSF grant DMR-0706020 and the 
Division of Materials Science and Engineering, U.S. DOE, under contract with 
UT-Battelle, LLC.  M.R. and A.M. are grateful to the Aspen Center for
Physics for its hospitality.
\end{acknowledgments}

\newpage

\appendix

\section*{Electronic physics auxiliary publication service for:
Quantum distillation: dynamical generation of low-entropy states of strongly 
correlated fermions in an optical lattice}

\subsection*{Expansion from a box trap: dependence on the initial filling $n_{\mathrm{init}}$}

\begin{figure}
\includegraphics[width=0.48\textwidth]{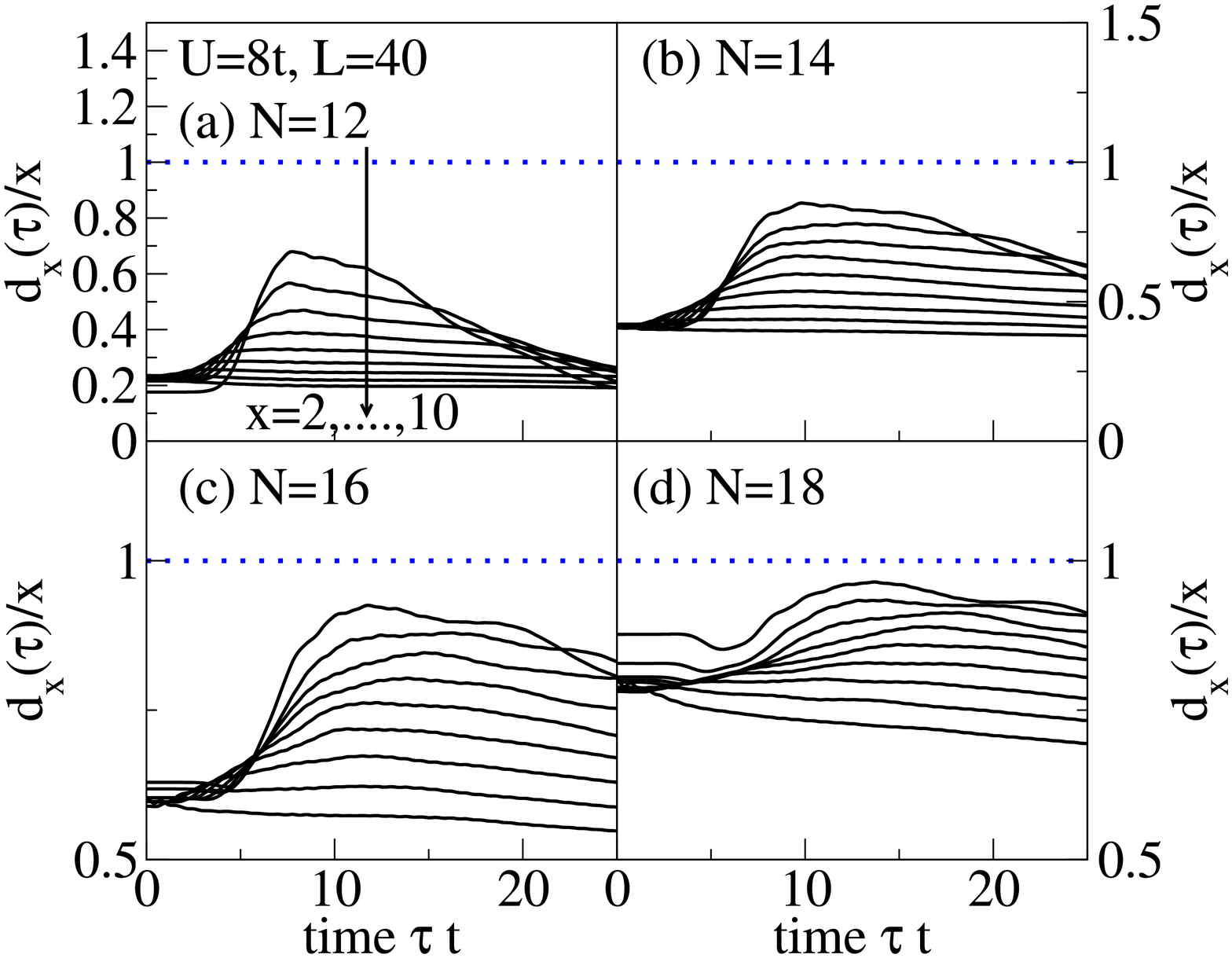}
\caption{(Color online)  Average double occupancy 
$d_{x}(\tau)/x$  for $x=2,\dots,10$ ($d_{x}(\tau)=\sum_{i=0}^{x} \langle \tilde d_i(\tau)\rangle$)
and $U=8t$: (a) $n_{\mathrm{init}}=1.2$; (b) $n_{\mathrm{init}}=1.4$  (c) $n_{\mathrm{init}}=1.6$;  (d) 
  $n_{\mathrm{init}}=1.8$. (time step $0.01/t$, discarded weight $10^{-8}$).
}
\label{fig:U8}
\end{figure}

\begin{figure}
\includegraphics[width=0.48\textwidth]{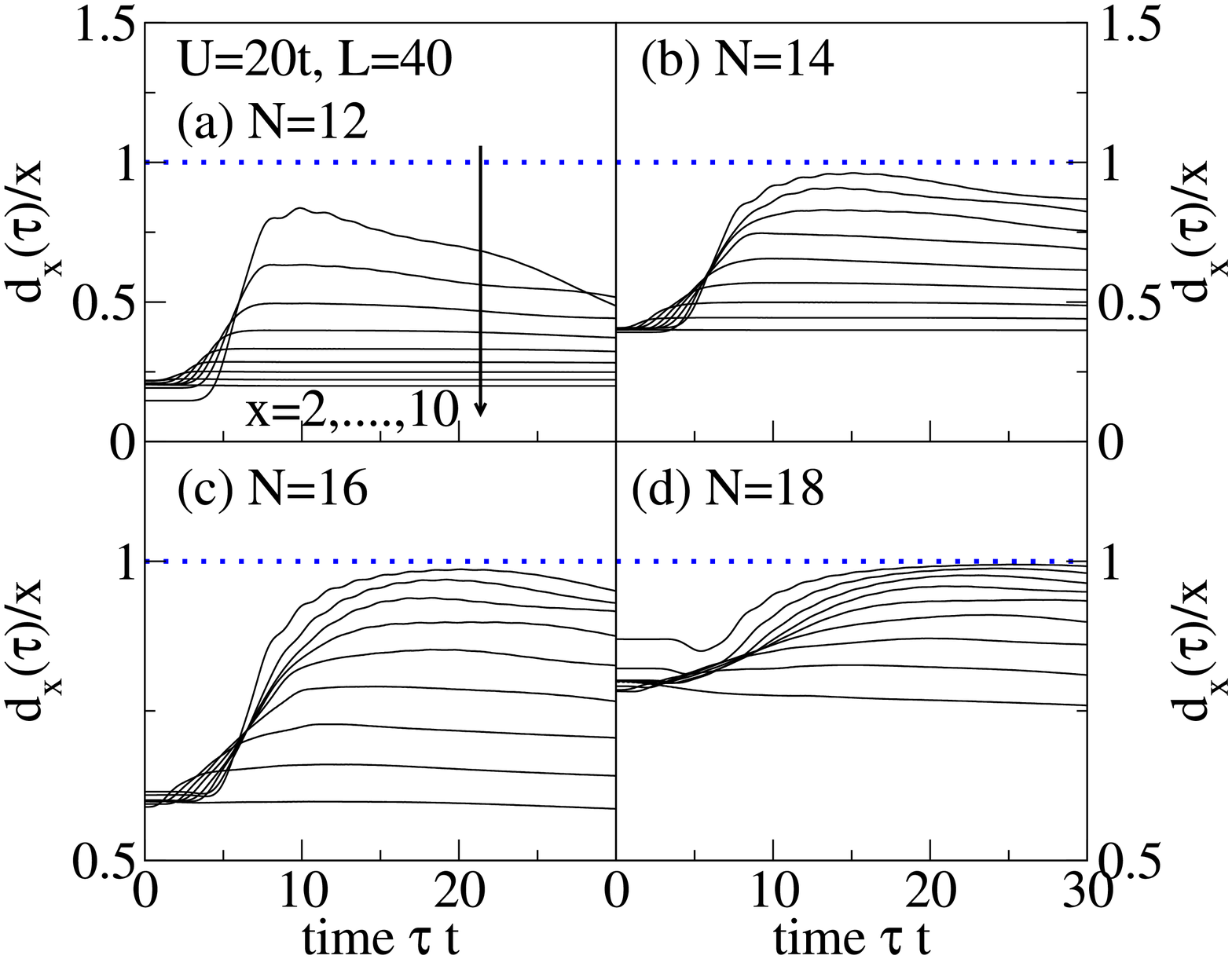}
\caption{(Color online)  Average double occupancy 
$d_{x}(\tau)/x$  for $x=2,\dots,10$ ($d_{x}(\tau)=\sum_{i=0}^{x} \langle \tilde d_i(\tau)\rangle$)
and $U=20t$: (a) $n_{\mathrm{init}}=1.2$; (b) $n_{\mathrm{init}}=1.4$  (c) $n_{\mathrm{init}}=1.6$;  (d) 
 $n_{\mathrm{init}}=1.8$. (time step $0.01/t$, discarded weight $10^{-8}$).
}
\label{fig:U20}
\end{figure}

\begin{figure}
\includegraphics[width=0.48\textwidth]{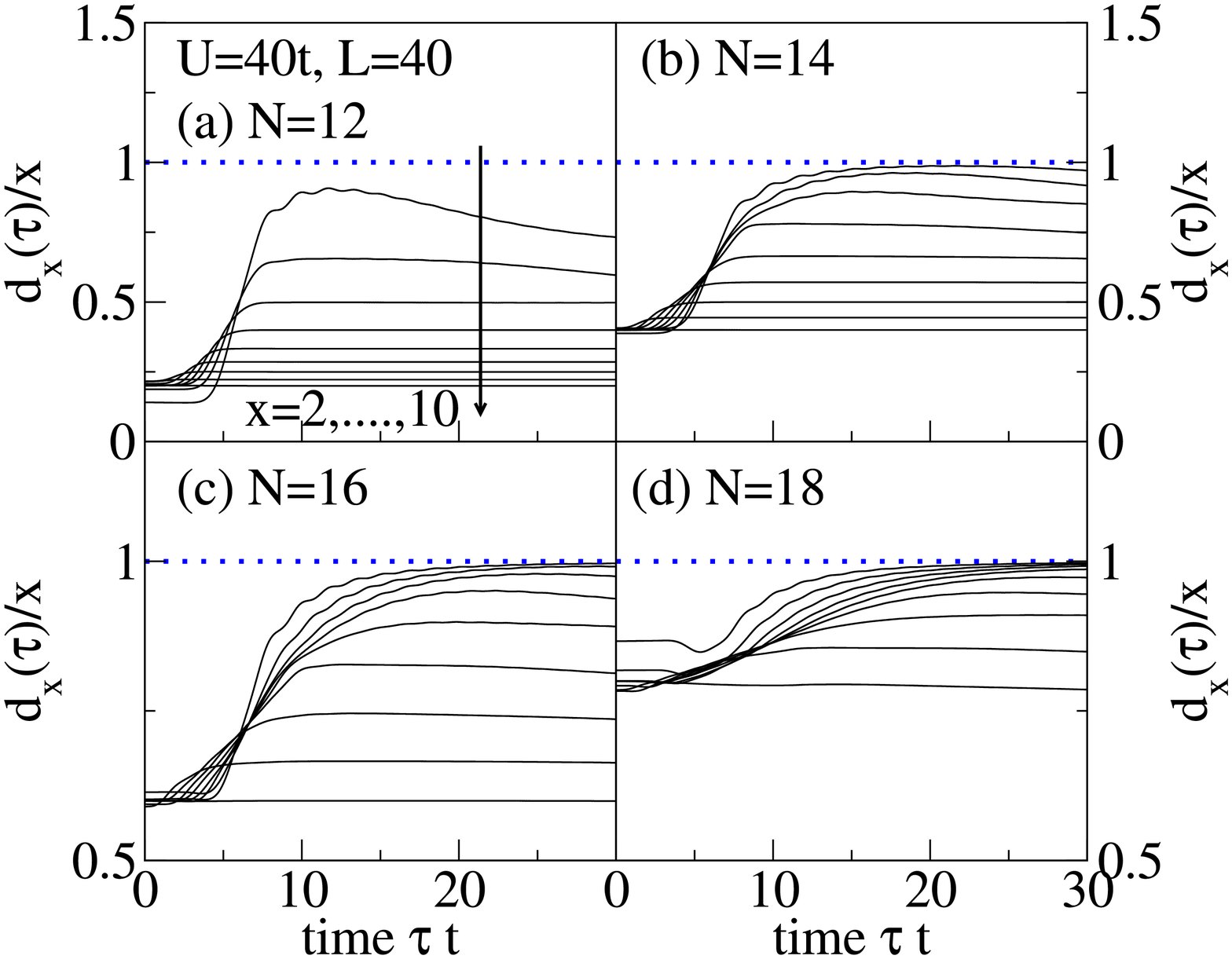}
\caption{(Color online)  Average double occupancy 
$d_{x}(\tau)/x$  for $x=2,\dots,10$ ($d_{x}(\tau)=\sum_{i=0}^{x} \langle \tilde d_i(\tau)\rangle$)
and $U=40t$: (a) $n_{\mathrm{init}}=1.2$; (b) $n_{\mathrm{init}}=1.4$  (c) $n_{\mathrm{init}}=1.6$;  (d) 
  $n_{\mathrm{init}}=1.8$. (time step $0.01/t$, discarded weight $10^{-9}$).
}
\label{fig:U40}
\end{figure}

\begin{figure}
\includegraphics[width=0.48\textwidth]{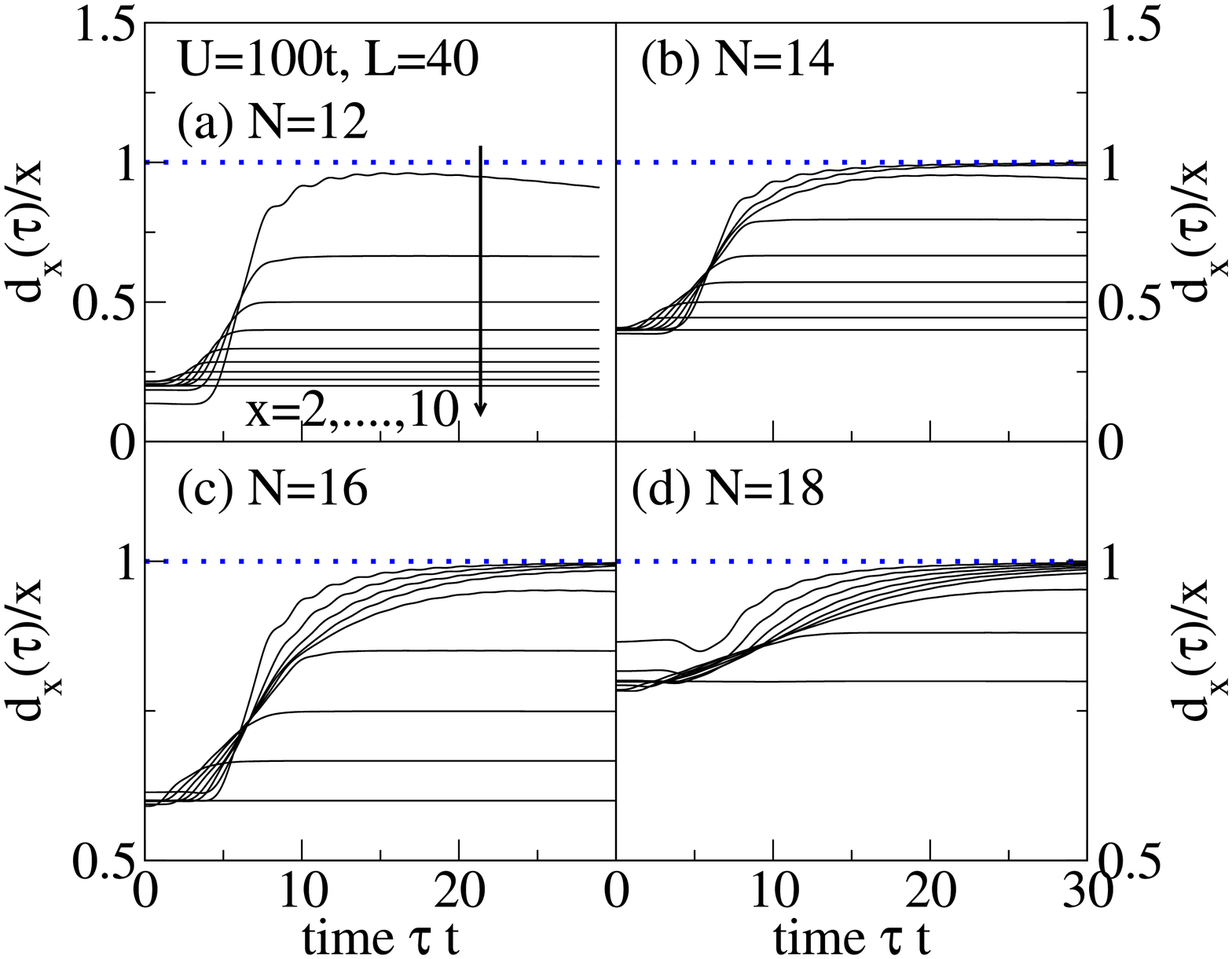}
\caption{(Color online)  Average double occupancy 
$d_{x}(\tau)/x$  for $x=2,\dots,10$ ($d_{x}(\tau)=\sum_{i=0}^{x} \langle \tilde d_i(\tau)\rangle$)
and $U=100t$: (a) $n_{\mathrm{init}}=1.2$; (b) $n_{\mathrm{init}}=1.4$  (c) $n_{\mathrm{init}}=1.6$;  (d) 
  $n_{\mathrm{init}}=1.8$. (time step $0.01/t$, discarded weight $10^{-9}$).
}
\label{fig:U100}
\end{figure}

In the main text, we have mostly discussed and presented results for an initial filling of $n_{\mathrm{init}}=1.8$ with $N=18$.
Here we present additional material for $n_{\mathrm{init}}=1.2,1.4,1.6$ (and, for comparison, the $n_{\mathrm{init}}=1.8$ data from the main
text)
to elucidate the dependence of the quantum distillation on $n_{\mathrm{init}}$. 
Such data are presented for $U/t=8,20,40,100$ in Figs.~\ref{fig:U8}-\ref{fig:U100}, respectively.

We find that 
for all $n_{\mathrm{init}}>1$ and all values of $U$ considered here, quantum distillation takes place.
We distinguish between the quantum distillation 
 in a {\it weak} sense, namely for some 
sites  $\langle n_i(\tau)\rangle>\langle n_i(\tau=0)\rangle$. Such behavior is seen in the case of $U=8t$
(Fig.~\ref{fig:U8}). The more pronounced effect of the
formation of approximate Fock states, {\it i.e.}, $\langle n_i(\tau)\rangle\approx 2$, is observed 
for large $n_{\mathrm{init}}$ and large $U$ (see Figs.~\ref{fig:U20}--\ref{fig:U100}). 

\subsection*{Expansion from a box trap: dependence on the particle number at a fixed initial filling $n_{\mathrm{init}}$}

\begin{figure}
\includegraphics[width=0.48\textwidth]{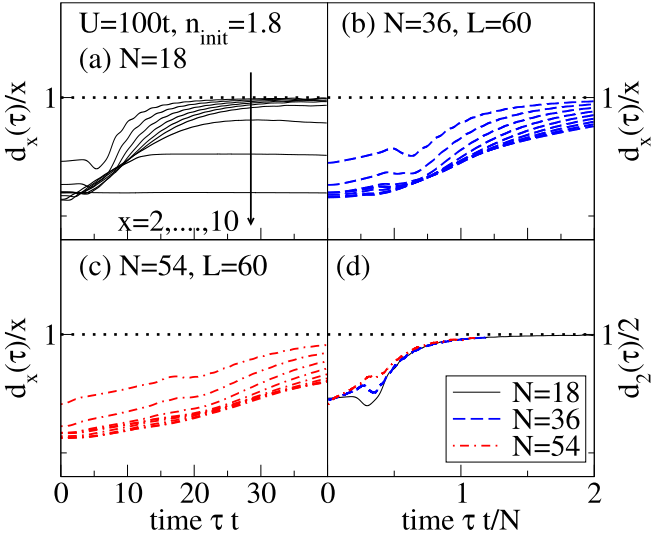}
\caption{(Color online)  Dependence of the average double occupancy 
$d_{x}(\tau)/x$  on the number of particles at a fixed filling of  $n_{\mathrm{init}}=1.8$ ($U=100t$): 
(a) $N=18$, $L=40$ ($\delta \tau\,t=0.01$, $\delta \rho=10^{-9}$); 
(b) $N=36$, $L=60$ ($\delta \tau\, t=0.01$, $\delta \rho=10^{-7}$); (c) $N=36$, $L=60$ ($\delta \tau\,t=0.02$, $\delta \rho=10^{-7}$); 
(d) $d_2(\tau)/2$ vs. time $\tau/N$ ({\it i.e.}, rescaled on the number of particles). The same qualitative behavior is observed, independently
of $N$. The time for building up the quasi-Fock state is simply proportional to the total number of particles. Slight differences between the 
results for different $N$ at short times are due to differences in the initial state ({\it i.e.}, boundary and $N$-dependent oscillations in the
density profile).
}
\label{fig:Ndep}
\end{figure}

We have also studied the dependence on the number of particles 
$N$ at a fixed $n_{\mathrm{init}}$. The respective results are presented in Fig.~\ref{fig:Ndep} for $U=100t$ and $n_{\mathrm{init}}=1.8$.
Qualitatively, the
  spatial extension and stability of the approximate Fock states 
grows with $N$. The figure further shows [see Fig.~\ref{fig:Ndep}(d)] that by plotting the data vs $\tau/N$, $d_x/x$, calculated for different $N$, all results collapse onto essentially
the same curve. This illustrates that the time for the formation of the quasi-Fock state scales linearly with the number of particles (or the 
number of doublons in the initial state, respectively). Most importantly, the {\it qualitative} behavior of the quantum distillation process
is independent of $N$ at a fixed $n_{\mathrm{init}}$

For $n_{\mathrm{init}}\leq 1$, no quantum distillation takes place at any $U$, and the 
dynamics in that case was studied in Ref.\ \cite{hm08a}.

\subsection*{Radius of the double occupancies}

\begin{figure}
\includegraphics[width=0.48\textwidth]{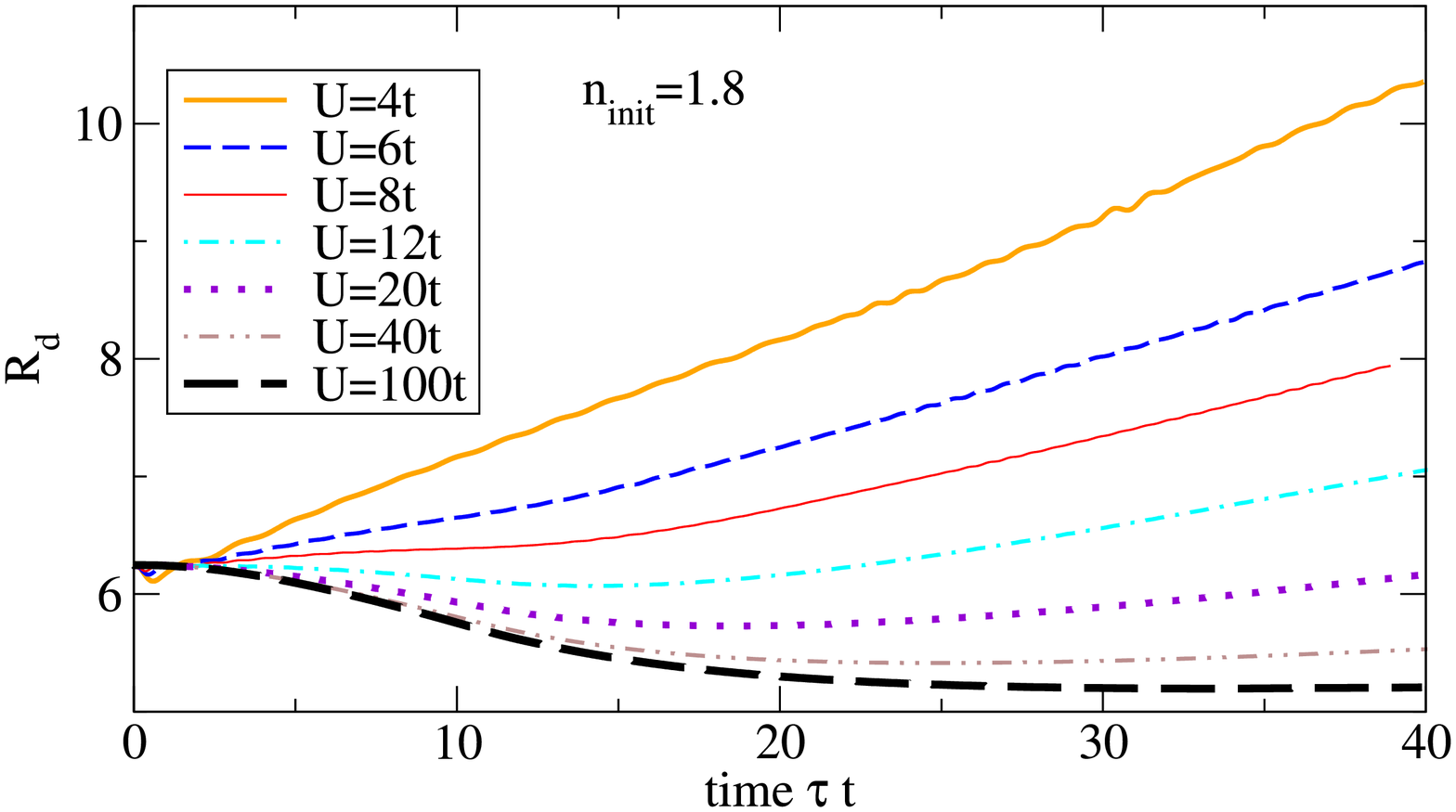}
\caption{(Color online) Radius $R_d$ of the double occupancies for the expansion from a box trap
with an initial density of $n=1.8$ and several $U/t=4,6,8,12,16,20,40,100$.
}
\label{fig:rd}
\end{figure}

The quantum distillation effect can be nicely illustrated by plotting the radius $R_d$ of the double occupancies:
\begin{equation}
R_d^2= \frac{1}{N_d}\sum_{i=1}^L i^2 \langle \tilde d_i \rangle 
\end{equation}
where $N_d= \sum_{i=1}^L \langle \tilde d_i\rangle $ is the total number of double occupancies.
This radius is shown in  Fig.\ref{fig:rd} for several values of $U$ and an initial density of
$n=1.8$. In this quantity, the effect is visible for $U\gtrsim 8t$.

\subsection*{Expansion from the harmonic trap: Reduction of the entanglement entropy}

\begin{figure}
\includegraphics[width=0.45\textwidth]{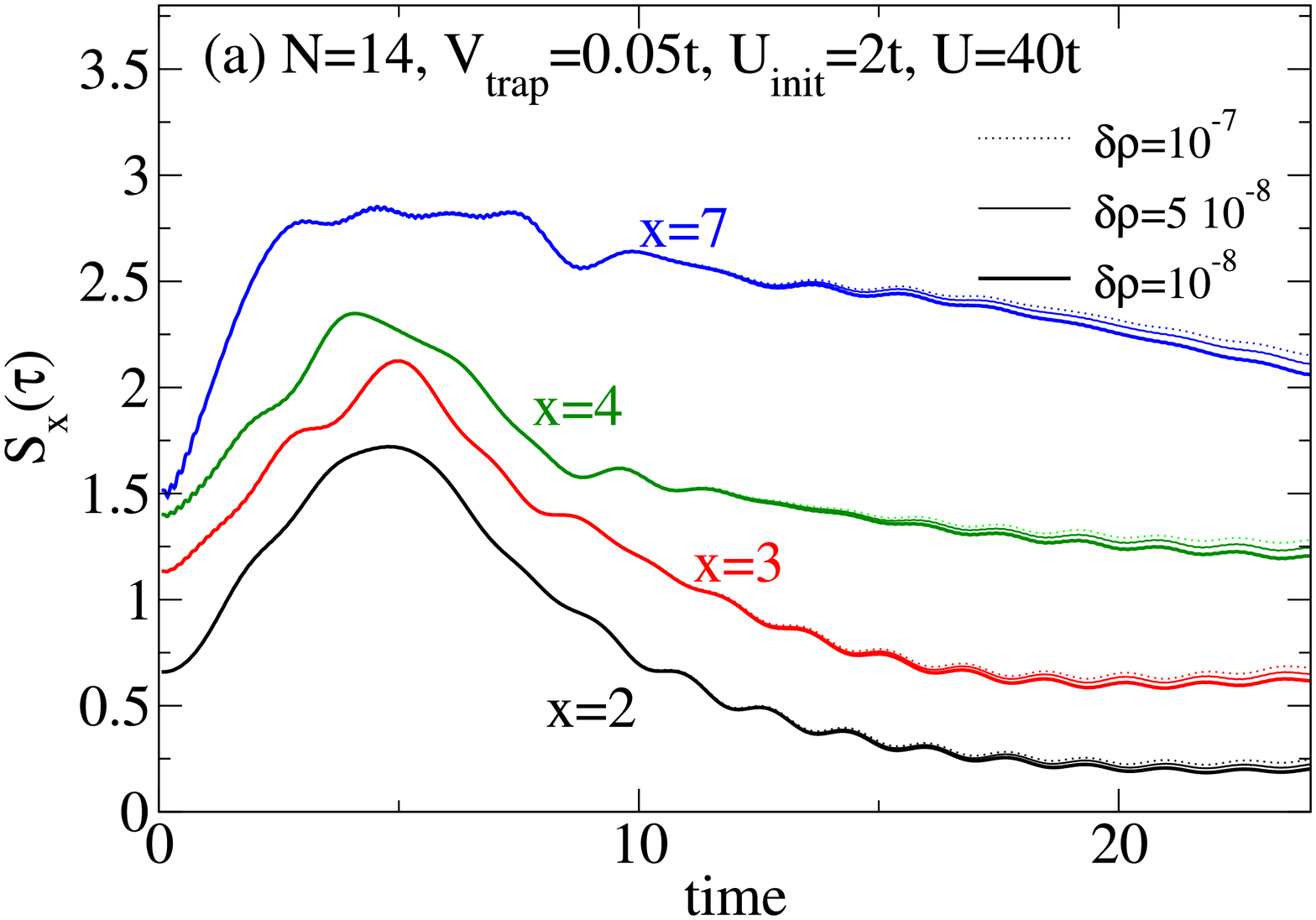}
\includegraphics[width=0.45\textwidth]{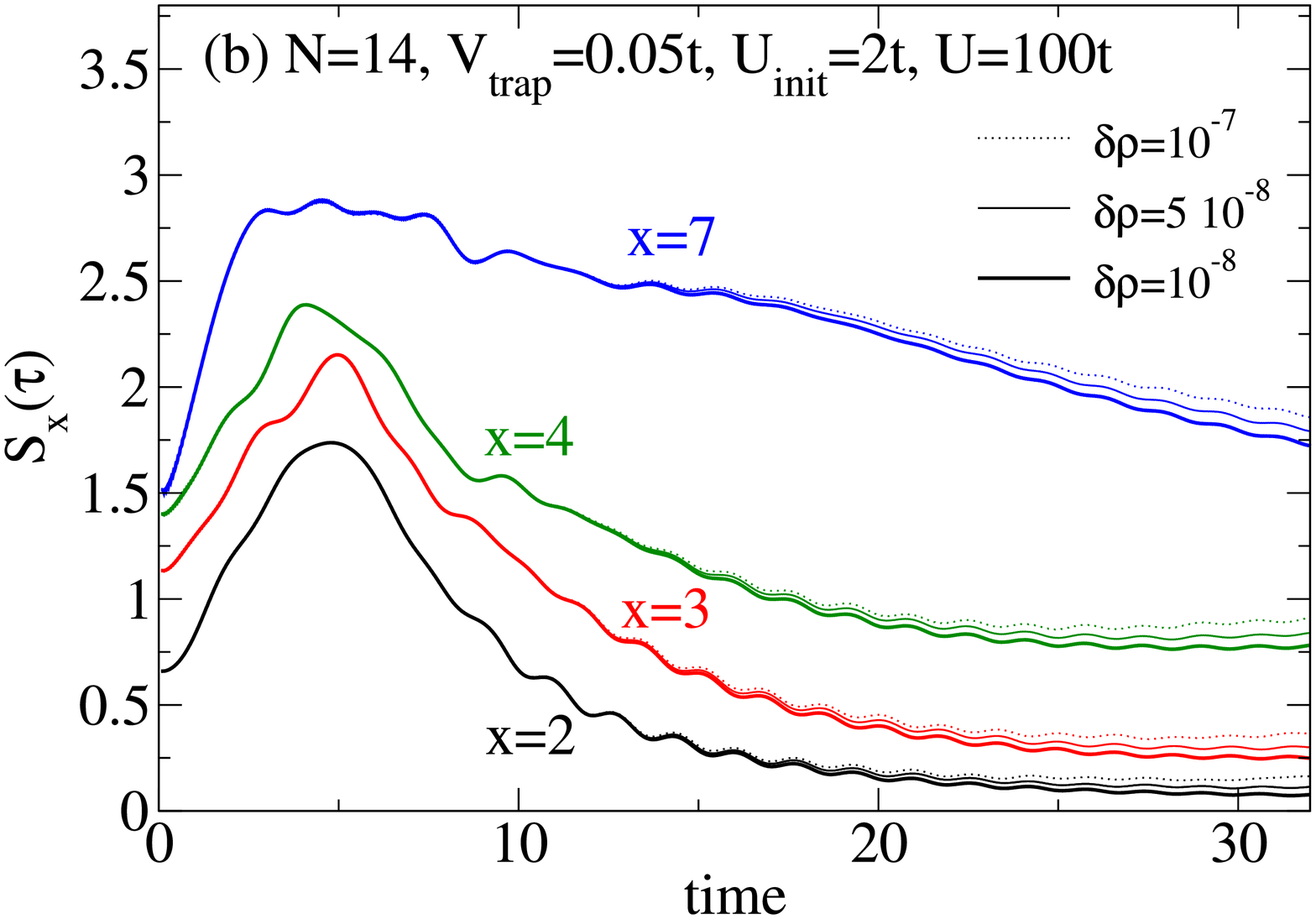}
\caption{(Color online)  Reduction of the entanglement entropy $S_x(\tau)$ for $x=2,3,4,7$ (bottom to top)
and the parameters of Fig.~4 of the main text ($V_{\mathrm{trap}}=0.05t$, $N=14$, $U_{\mathrm{init}}$=2t): (a) $U=40t$; (b) $U=100t$.
The runs were performed with a time step of $0.01/t$ and different discarded weights $\delta \rho$ (see the figure's legend).}
\label{fig:entropy_app}
\end{figure}

Here we provide results for the reduction of the entanglement entropy for the expansion from a harmonic trap
and the parameters of Fig. 4 of the main text ($V_{\mathrm{trap}}=0.05t$, $N=14$, $U_{\mathrm{init}}=2t$), yet for two values
of the interaction $U=40t$ and $U=100t$ after the quench. We have also performed runs with different discarded
weights $\delta \rho$ to check 
the quality of our results. The data presented in Fig.~\ref{fig:entropy_app}(b) suggest that the reduction of the entanglement
entropy is (for $x=2,3$ and $U=100t$) 
\begin{eqnarray}
S_{x=2}(\tau) &<& 0.1 \,S_{x=2}(\tau=0) \\\nonumber
S_{x=3}(\tau) &<&  0.21\, S_{x=3}(\tau=0)\,.
\end{eqnarray}
Note that we have not yet reached the minimum in the time-dependent reduction of $S_{x}(\tau)$. At  long times, the simulation
is slowed down due to the relatively large entanglement  of blocks with $x\gtrsim 7$. The figure unveils that the more accurate the simulation is
({\it i.e.}, the smaller the discarded weight), the better the  efficiency of the quantum distillation in reducing $S_x(\tau)$ is captured.

\begin{figure}[t]
\includegraphics[width=0.48\textwidth]{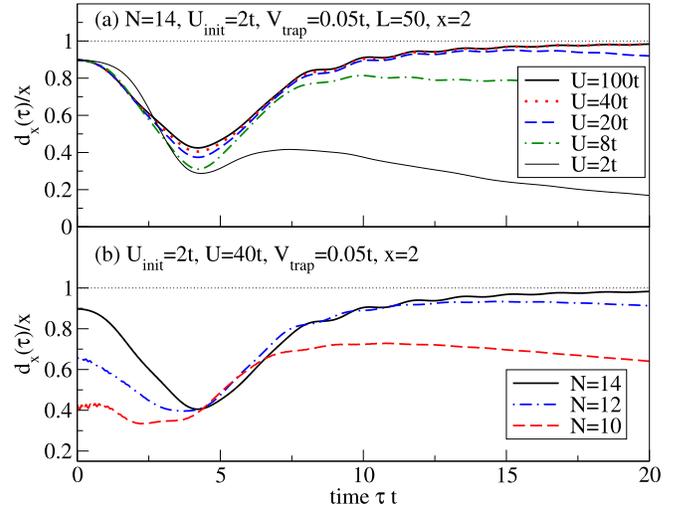}
\caption{(Color online) Average double occupancy $d_2/2$ for the set-up of Fig.~4 from the main text  ($V_{\mathrm{trap}}=0.05t$,  $U_{\mathrm{init}}=2t$)
and (a) $U/t=2,8,20,40,100$ ($N=14$) and (b) $N=10,12,14$ ($U=40t$).}
\label{fig:trap_N}
\end{figure}

The last figure, Fig.~\ref{fig:trap_N}, illustrates the dependence of the quantum distillation on first, $U$  [panel (a)] 
and second, the effective density $\tilde \rho=N\sqrt{V_{\mathrm{trap}}}$ [panel (b)].
We use the parameters of Fig.~4 from the main text, except for varying $U$ in Fig.~\ref{fig:trap_N}(a) and varying $N$  in  Fig.~\ref{fig:trap_N}(b). We plot the average double occupancy and we observe that similar
to the case of a box trap that was discussed in the main text, quantum distillation works as long as some doublons (signified by sites with $\langle n_i\rangle>1$ at $\tau=0$)
are present in the initial state. Also, as expected, the final $U$ needs to be larger than $4t$ to induce the formation of a metastable state with
doublons grouping together in the leftmost sites.

\clearpage

\clearpage

\begin{thebibliography}{10}
\bibitem{quench}
In experiments, $U/t$ should not be too large to avoid
the population of higher bands.
\end{thebibliography}
\end{document}